\documentstyle[epsf]{mn}

\newif\ifAMStwofonts

\ifoldfss
  \ifCUPmtlplainloaded \else
    \NewTextAlphabet{textbfit} {cmbxti10} {}
    \NewTextAlphabet{textbfss} {cmssbx10} {}
    \NewMathAlphabet{mathbfit} {cmbxti10} {} 
    \NewMathAlphabet{mathbfss} {cmssbx10} {} 
  \fi
  \ifAMStwofonts
    \ifCUPmtlplainloaded \else
      \NewSymbolFont{upmath} {eurm10}
      \NewSymbolFont{AMSa} {msam10}
      \NewMathSymbol{\upi}     {0}{upmath}{19}
      \NewMathSymbol{\umu}     {0}{upmath}{16}
      \NewMathSymbol{\upartial}{0}{upmath}{40}
      \NewMathSymbol{\leqslant}{3}{AMSa}{36}
      \NewMathSymbol{\geqslant}{3}{AMSa}{3E}

    \fi
  \fi
\fi 

\ifnfssone
  \newmathalphabet{\mathit}
  \addtoversion{normal}{\mathit}{cmr}{m}{it}
  \addtoversion{bold}{\mathit}{cmr}{bx}{it}
  \newmathalphabet{\mathbfit} 
  \addtoversion{normal}{\mathbfit}{cmr}{bx}{it}
  \addtoversion{bold}{\mathbfit}{cmr}{bx}{it}
  \newmathalphabet{\mathbfss} 
  \addtoversion{normal}{\mathbfss}{cmss}{bx}{n}
  \addtoversion{bold}{\mathbfss}{cmss}{bx}{n}
  \ifAMStwofonts
    \ifCUPmtlplainloaded \else
      \UseAMStwoboldmath
      \makeatletter
      \new@mathgroup\upmath@group
      \define@mathgroup\mv@normal\upmath@group{eur}{m}{n}
      \define@mathgroup\mv@bold\upmath@group{eur}{b}{n}
      \edef\UPM{\hexnumber\upmath@group}
      \new@mathgroup\amsa@group
      \define@mathgroup\mv@normal\amsa@group{msa}{m}{n}
      \define@mathgroup\mv@bold\amsa@group{msa}{m}{n}
      \edef\AMSa{\hexnumber\amsa@group}
      \makeatother
      \mathchardef\upi="0\UPM19
      \mathchardef\umu="0\UPM16
      \mathchardef\upartial="0\UPM40
      \mathchardef\leqslant="3\AMSa36
      \mathchardef\geqslant="3\AMSa3E
    \fi
  \fi
\fi 

\ifnfsstwo
  \DeclareMathAlphabet{\mathbfit}{OT1}{cmr}{bx}{it}
  \SetMathAlphabet\mathbfit{bold}{OT1}{cmr}{bx}{it}
  \DeclareMathAlphabet{\mathbfss}{OT1}{cmss}{bx}{n}
  \SetMathAlphabet\mathbfss{bold}{OT1}{cmss}{bx}{n}
  \ifAMStwofonts
    \ifCUPmtlplainloaded \else
      \DeclareSymbolFont{UPM}{U}{eur}{m}{n}
      \SetSymbolFont{UPM}{bold}{U}{eur}{b}{n}
      \DeclareSymbolFont{AMSa}{U}{msa}{m}{n}
      \DeclareMathSymbol{\upi}{0}{UPM}{"19}
      \DeclareMathSymbol{\umu}{0}{UPM}{"16}
      \DeclareMathSymbol{\upartial}{0}{UPM}{"40}
      \DeclareMathSymbol{\leqslant}{3}{AMSa}{"36}
      \DeclareMathSymbol{\geqslant}{3}{AMSa}{"3E}
    \fi
  \fi
\fi 

\ifCUPmtlplainloaded \else
  \ifAMStwofonts \else 
    \def\upi{\pi}
    \def\umu{\mu}
    \def\upartial{\partial}
  \fi
\fi

\title[Spectral variation in GX 1+4]
  {Spectral variation in the X-ray pulsar GX 1+4 during a low-flux episode}
\author[D.K. Galloway et al.]
  {D.K. Galloway,$^{1,2}$ A.B. Giles,$^{1,3,4}$ J.G. Greenhill,$^{1}$
  and M.C. Storey$^{2}$ \\
  $^1$ School of Mathematics and Physics,
       University of Tasmania,
       GPO Box 252-21, Hobart,
       Tasmania 7001, Australia \\
  $^2$ Special Research Centre for Theoretical Astrophysics,
       School of Physics, University of Sydney,
       NSW, Australia, 2006 \\
  $^3$ Laboratory for High Energy Astrophysics,
       Goddard Space Flight Center,
       Greenbelt, MD 20771, USA\\
  $^4$ Universities Space Research Association}

\pagerange{\pageref{firstpage}--\pageref{lastpage}}
\pubyear{1998}

\begin{document}

\maketitle

\label{firstpage}


\begin{abstract}
The X-ray pulsar GX 1+4 was observed with the {\it RXTE} satellite for a total
of 51~ks between 1996 July 19 - 21. During this period the flux
decreased smoothly from an initial mean level of $\approx 6\times
10^{36}\,{\rm erg\,s^{-1}}$ to a minimum of $\approx 4\times 10^{35}\,{\rm
erg\,s^{-1}}$ (2-60~keV, assuming a source distance of 10~kpc) before
partially recovering towards the initial level at the end of the observation.

BATSE pulse timing measurements indicate that a torque reversal took place
approximately 10~d after this observation. Both the mean pulse profile and
the photon spectrum varied significantly. The observed variation in the
source may provide important clues as to the mechanism of torque reversals.

The single best-fitting spectral model was based on a component originating
from thermal photons with $kT_0 \approx 1$~keV Comptonised by a plasma of
temperature $kT \approx 7$~keV. Both the flux modulation with phase during
the brightest interval and the evolution of the mean spectra over the course
of the observation are consistent with variations in this model component;
with, in addition, a doubling of the column density $n_H$ contributing to the
mean spectral change.

A strong flare of duration $\la 50$~s was observed during the interval
of minimum flux, with the peak flux $\approx 20$ times the mean level.
Although beaming effects are likely to mask the true variation in ${\dot M}$
thought to give rise to the flare, the timing of a modest increase in flux
prior to the flare is consistent with dual episodes of accretion resulting
from successive orbits of a locally dense patch of matter in the accretion
disc.

\end{abstract}

\begin{keywords}
binaries: symbiotic -- pulsars: individual (GX 1+4) -- X-rays: stars
\end{keywords}

\section{Introduction}

The study of X-ray pulsars has been an area of active research for almost 30
years.  In spite of this there remain some significant shortfalls in the
understanding of these objects. An example is the persistent pulsar GX 1+4.
At the time of its discovery \cite{lew71} it was one of the brightest
objects in the X-ray sky. The companion to the neutron star is an M6 giant
\cite{dav77}.  GX 1+4 is the only known X-ray pulsar in a symbiotic system.
Measurements of the average spin-up rate during the 1970s gave the largest
value recorded for any pulsar (or in fact any astronomical object) at
$\approx 2$ per cent per year. Inexplicably, the average spin-up trend
reversed around 1983, switching to spin-down at approximately the same rate.
Since that reversal a number of changes in the sign of the torque (as
inferred from the rate of change of the pulsar spin period) have been
observed \cite{chak97b}.  Several estimates (Beurle et al. 1984, Dotani et
al. 1989, Greenhill et al.  1993, Cui 1997) indicate a neutron star surface
magnetic field strength of $2-3 \times 10^{13}$~G.

The X-ray flux from the source is extremely variable on time-scales of seconds
to decades. Two principal flux states have been observed, a `high' state which
persisted during the spin-up period of the 1970s, and a `low' state since.
Although the mean flux has been increasing steadily during the current `low'
state it has not yet returned to the level of the 1970s.  Superimposed on
these long-term variations are smooth changes in the flux on time-scales of
order hours to days. On the shortest time-scales the periodic variation due to
the neutron star's rotation period at around 2 min is observed.


Compared to other accretion-powered X-ray pulsars, GX 1+4 has an atypically
hard spectrum extending out well past 100~keV \cite{fro89}.  Historically the
spectrum has been fitted with thermal bremsstrahlung or power law models;
more recent observations with improved spectral resolution generally favour
a power law model with exponential cutoff.
Typical values for the cutoff power law model parameters are photon index
$\alpha = 1.1-2.5$; cutoff energy $5-18$~keV; $e$-folding energy
$11-26$~keV.  For any spectral model covering the range 1-10~keV, it is also
necessary to include a gaussian component representing iron line emission at
$\approx 6.4$~keV, and a term to account for the effects of photoelectric
absorption by cold gas along the line of sight with hydrogen column density
in the range $n_H = (4-140) \times 10^{22}\,{\rm cm^{-2}}$.
The source spectrum and in particular the column density $n_H$ have
previously exhibited significant variability on time-scales as short as a
day \cite{beck76}.  Measurements of spectral variation with phase are few;
one example of pulse-phase spectroscopy was undertaken with data from the
{\it Ginga} satellite from 1987 and 1988 \cite{dot89}.  Only the column
density and the iron line centre energy were allowed to vary with phase in
the spectral fits, and no significant variation was observed.

The Ghosh and Lamb \shortcite{gl79} model predicts a correlation between
torque and mass transfer rate (and hence luminosity) for accretion-driven
X-ray sources. For most sources it is difficult to test such a relationship
since the range of luminosities at which they are observed is limited.
However the correlation between torque and luminosity has been confirmed, at
least approximately, for three transient sources using data from the Burst
and Transient Source Experiment (BATSE) aboard the Compton Gamma Ray
Observatory ({\it CGRO}) and {\it EXOSAT} (Reynolds et al. 1996, Bildsten et
al. 1997). 

The situation for persistent pulsars is, however, less straightforward. The
BATSE data have demonstrated that in general the torque is in fact {\it
uncorrelated} with luminosity in these sources. The spin-up or spin-down
rate can remain almost constant over intervals (referred to in this paper as
a `constant torque state' or just `torque state') which are long compared to
other characteristic time-scales of the system, even when the luminosity
varies by several orders of magnitude over that time. 
Transitions between these states of constant torque can be abrupt, with
time-scales of $< 1$ d when the two torque values have the same sign;
alternatively when switching from spin-up to spin-down (or vice-versa) the
switch generally occurs smoothly over a period of $10 - 50$ d.

It is possible that there remains some connection between the torque and
luminosity, since at times the torque measured for GX 1+4 has been {\it
anticorrelated} with luminosity \cite{chak97b}. This behaviour has not been
observed in other pulsars. One important caveat regarding the BATSE
measurements is that the instrument can only measure pulsed flux.  Systematic
variations in pulse profiles or pulse fraction could introduce significant
aliasing to the flux data, hence masking the true relationship between
bolometric flux and torque.  Given that pulse profile shape and torque state
have shown evidence for correlation in GX 1+4 \cite{gre98} this could
potentially be an important effect. 

In this paper we present results from spectral analysis of data obtained
from GX 1+4 during 1996 using the Rossi X-ray Timing Explorer satellite
({\it RXTE}; Giles et al. 1995). A companion paper \cite{gil99} contains
detailed analysis of pulse arrival times and pulse profile changes.

\section{Observations}

The source was observed with {\it RXTE} between 1996 July 19 16:47 UT and
1996 July 21 02:39 UT.  Several interruptions were made during that time as
a consequence of previously scheduled monitoring of other sources.  After
screening the data to avoid periods contaminated by Earth occultations, the
passage of the satellite through the South Atlantic Anomaly (SAA), and
periods of unstable pointing, the total on-source duration was $51$ ks.
{\em RXTE} consists of three instruments, the proportional counter array
(PCA) covering the energy range 2-60~keV, the high-energy X-ray timing
experiment (HEXTE) covering 16-250~keV, and the all-sky monitor (ASM) which
spans 2-10~keV. Pointed observations are performed using the PCA and HEXTE
instruments, while the ASM regularly scans the entire visible sky.

\begin{figure}
  \epsfxsize=9.0cm
  \epsfbox{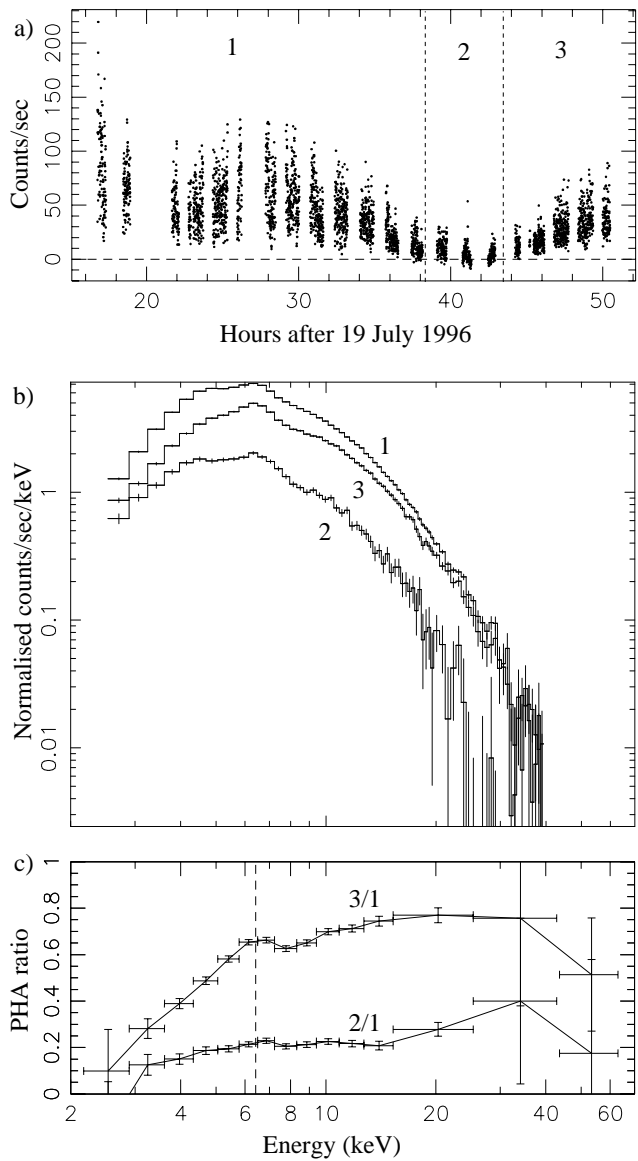}
  \caption{
a) Background-subtracted PCA countrate for GX 1+4 over the course of the
observation, showing the division between intervals 1, 2 and 3 (see text).
The binsize is 16 s.  b) Mean PCA spectra between 2.2 and 40.0~keV during
each interval.  The data mode used is Standard-2, with 128 channels over the
total range of the instrument.  c) Ratio of mean spectra in intervals 2 and
3 to that of interval 1, labelled 2/1 and 3/1 respectively. The dotted line
indicates the energy of the Fe line. Spectra were obtained from the PCA
except for the point covering the highest energy range, which was calculated
using HEXTE data.
\label{fig1}}
\end{figure}

The background-subtracted total PCA count rate for 3 of the five
proportional counter units (PCUs) comprising the PCA is shown in Fig.
\ref{fig1}a. The other two PCUs were only active briefly at the beginning of
the observation so those data are not included in the analysis. The
phase-averaged PCA count rate was initially low at $\approx 80\,{\rm
count\,s^{-1}}$. This corresponds to a flux of $\approx 6 \times
10^{36}\,{\rm erg\,s^{-1}}$ in the 2-60~keV energy range, using the spectral
model discussed in section \ref{spec} and assuming a source distance of
10~kpc. Throughout this paper we shall use this value as the source distance
unless otherwise specified; the actual distance is thought to be in the
range 3-15 kpc \cite{chak97}.  During the course of the observation the
count rate decreased to a minimum of $\approx 5\,{\rm count\,s^{-1}}$,
corresponding to a flux of $\approx 4 \times 10^{35}\,{\rm erg\,s^{-1}}$
before partially recovering towards the end.  The count rates are unusually
low for this source, with other observations giving significantly higher
rates; for example $\approx 320\,{\rm count\,s^{-1}}$ and $230\,{\rm
count\,s^{-1}}$ (equivalent rates for 3 PCU's) in February 1996 and January
1997 respectively.
At times the background-subtracted countrate during interval 2 drops
significantly below zero. This is a consequence of the low source to
background signal ratio (around 1:10) during this interval coupled with
statistical variations in the binned countrate values.

The observation is divided into three intervals on the basis of the mean
flux (Fig. \ref{fig1}a). Interval 1 covers the start of the observation to
just before the flux minimum. Interval 2 spans the period of minimum flux,
during which time the flux was $\la 30\,{\rm count\,s^{-1}}$ apart from
$\approx 10$~s during a flare (see section \ref{flare}). Interval 3 covers
the remaining portion of the observation, during which the mean flux
increased steadily.


Accompanying the changes in flux were significant variations in pulse profile
and spectral shape. Historically GX 1+4 has shown evidence of a correlation
between torque state and pulse profile shape \cite{gre98}. Throughout the
period of spin-up during the 1970s pulse profiles were typically brighter at
the trailing edge with respect to the primary minimum; e.g. Doty, Hoffman and
Lewin \shortcite{dot81}.  Since then, measured pulse profiles have instead
usually been leading-edge bright, with less pronounced asymmetry; e.g.
Greenhill et al.  \shortcite{gre93}.  During interval 1 the pulse profile was
observed to be leading-edge bright, similar to other observations since the
1980s. Pulsations all but ceased during interval 2, and in interval 3 the
shape of the profile had changed dramatically and resembled the trailing-edge
bright profiles typically observed during the 1970s \cite{gil99}.

The count rate spectra taken during each interval are shown in Fig.
\ref{fig1}b.  The overall spectral shape changed significantly over the course
of the observation, with the spectrum becoming harder in intervals 2 and 3
compared to interval 1.  The iron fluorescence line at around 6.4~keV appears
more prominent in the second and third intervals. Iron line enhancement
during intervals 2 and 3 is also apparent in the spectral ratios, Fig.
\ref{fig1}c. These ratios were calculated by subtracting the background
spectrum (including a component to account for the emission from the galactic
plane; see section \ref{spec}) from the source spectrum for each interval
and dividing the resulting spectra for intervals 2 and 3 by that of interval
1.  Because the countrate drops off steeply above 10~keV the spectral bins
must be made correspondingly larger to achieve a reliable ratio. The
datapoint in the highest energy band for each curve was obtained from HEXTE
data, while the lower energy ratios are calculated from PCA data. The
decrease in flux observed from interval 1 to 2 and 3 becomes more pronounced
at energies below 6~keV. Above 15~keV the spectral ratios are almost
constant with energy.

\section{The spectral model fits}
\label{spec}

Instrumental background from cosmic ray interactions and as a result of
passages close to the SAA are estimated using the {\sc pcabackest} software,
provided by the {\it RXTE} Guest Observer Facility (GOF). Due to the
proximity of the source to the galactic plane, an additional component which
takes into account the so-called `galactic ridge' emission must be
included in any spectral model.  The model for this component used in our
fits is identical to that fitted to survey data from this region
\cite{val98} with normalisations and abundance fitted to spectra taken
during slews to and from the source during this observation.  A secondary
instrumental effect which must be taken into account to obtain the lowest
possible residuals in the model fit is a consequence of the Xenon L-edge
absorption feature. This feature is modelled in our spectrum by a
multiplicative edge model component with energy fixed at 4.83~keV.

Candidate spectral models were tested by fitting to the count rate spectrum to
minimise $\chi^{2}$ using the {\sc xspec} spectral fitting package version 10
\cite{xspec}. In general each model takes the form of one (or more) continuum
components with a gaussian component necessary to simulate the iron line
emission, and a multiplicative component to account for absorption by cold
matter along the line of sight. With a lower than normal count rate for the
source during this observation, the primary source of error is the Poisson
statistics within each energy bin rather than any instrumental uncertainty.
Hence the systematic error in {\sc xspec} was set to zero for the model
fits.

Thermal bremsstrahlung and power law continuum components both resulted in
formally unacceptable values of $\chi^{2}$ for the interval 1 mean spectrum.
Some improvement was found by fitting with various forms of power law,
including a power law with exponential cutoff and a broken power law.
Nevertheless, each of these models gave unacceptable values of
reduced-$\chi^{2}$: 2.93 and 1.94 respectively.  An acceptable fit was
obtained using an analytic model based on Comptonisation of a thermal
spectrum by hot plasma close to the source (`{\tt compTT}' in {\sc xspec};
Titarchuk 1994), with reduced-$\chi^{2}$ for the interval 1 spectra of 1.10.
The Comptonisation model has thus been chosen for all spectral fitting
reported in this paper.  Model parameters for spectral fits from intervals 1
and 3 are shown in Table \ref{tab1}.

Fitting this model to the interval 2 mean spectra resulted in an acceptable
$\chi^{2}_{\nu}$ fit statistic of 0.7943, but with very wide confidence
limits for the fit parameters. No improvement in the confidence intervals is
obtained by freezing selected parameters to the mean values for the entire
observation (e.g. $T_{\rm 0}$). The fit parameters are effectively
unconstrained and as such cannot be relied upon as a measure of the source
conditions.  Additionally, the interval 2 spectra alone do not permit an
unambiguous choice of spectral model.  We cannot distinguish between cutoff
powerlaw, broken powerlaw, and Comptonisation spectral models during this
interval on the basis of $\chi^2_{\nu}$. A comparable fit can even be
obtained using a model consisting of two blackbody emission components, with
fitted temperatures
$1.4_{1.2}^{1.6}$~keV and $6.0_{4.5}^{7.6}$~keV ($\chi^2_{\nu} = 0.81$;
see section \ref{flare}).  Consequently we restrict the discussion of the
mean spectral fitting results to those from intervals 1 and 3.

The Comptonisation model implementation in {\sc xspec} offers a geometry
switch which affects the fitted value of the optical depth $\tau_P$.  The
switch can be set to model either disc or spherical geometries.  As we will
argue in section \ref{discussion}, neither of these are strictly appropriate
for the present situation.  Consequently we performed all the fitting using
the disc geometry, but note that fitted values of $\tau_P$ with the
spherical geometry will be approximately twice as large. The fitted values
of $\tau_P$ should provide an adequate comparative measure of the degree of
Comptonisation between different spectra.

\begin{table*}
\begin{minipage}{150mm}
  \caption{ Fit parameters for intervals 1 and 3 using the best-fitting model
    based on Comptonisation of soft photons by hot plasma. $kT_0$ is the
    temperature of the thermal input spectrum, $kT$ and $\tau_P$ are the
    temperature and optical depth respectively of the scattering plasma, and
    $A_{\rm C}$ is the normalisation parameter for the Comptonised model 
    component.
    The model also incorporates a gaussian component representing iron line
    emission, with $E_{\rm Fe}$ the line centre energy, $\sigma$ the line
    width, $A_{\rm Fe}$ the normalisation and EW the equivalent width. Both
    these components are attenuated by photoelectric absorption by cold
    matter along the line of sight, with column density $n_H$. Data used is
    from PCA mode Standard-2. Confidence intervals are 90 per cent; fit
    statistic is reduced-$\chi^2$ ($\chi^2_{\nu}$).
    \label{tab1} }
 \renewcommand{\arraystretch}{1.5}
\begin{center}
 \begin{tabular}{lccc}
  Parameter & Interval 1 & Interval 3 \\
  \hline 
$n_{H}$ ($ \times 10^{22}\,{\rm cm}^{-2}$) & $13.6^{14.4}_{12.9}$ & $28.8^{32.3}_{25.0}$ \\
$kT_0$ (keV) & $1.18^{1.21}_{1.15}$ & $1.00^{1.13}_{0.88}$ \\
$kT$ (keV) & $7.77^{8.53}_{7.19}$ & $8.61^{14.1}_{6.93}$ \\
$\tau_P$ & $3.26^{3.48}_{3.02}$ & $2.90^{3.51}_{1.97}$ \\
$A_{\rm C}$ ($ {\rm photons\,cm^{-2}\,s^{-1}\,keV^{-1}}$) & $(5.11^{5.56}_{4.64} ) \times 10^{-3}$ & $(4.42^{4.98}_{3.51} ) \times 10^{-3}$ \\
$E_{\rm Fe}$ (keV) & $6.406^{6.450}_{6.357}$ & $6.37^{6.46}_{6.20}$ \\
$\sigma$ & $0.32^{0.400}_{0.23}$ & $0.37^{0.59}_{0.20}$ \\
$A_{\rm Fe}$ ($ {\rm photons\,cm^{-2}\,s^{-1}\,keV^{-1}}$) & $(4.23^{4.93}_{3.74} ) \times 10^{-4}$ & $(5.0^{8.3}_{3.8} ) \times 10^{-4}$ \\
EW (keV)& $0.18^{0.21}_{0.16}$ & $0.24^{0.36}_{0.19}$ \\
  \hline 
$\chi_{\nu}^{2}$ & 1.105 (64 d.o.f) & 0.7251 (64 d.o.f) \\
 \end{tabular}
\end{center}
\end{minipage}
\end{table*}

The increase in line-of-sight absorption following interval 1, suggested
initially by the spectral ratios (Fig. \ref{fig1}c), is further supported by
the model fits. The fitted column density $n_H$ more than doubles between
interval 1 and 3. Spectral fits to each uninterrupted `burst' of data (see
Fig. \ref{fig1}a) indicate that the increase took place smoothly over
approximately 10 hours, although significant variations in the fitted $n_H$
values are observed on timescales as short as 2~h.  {\it BeppoSAX }
satellite observations indicate that $n_H$ may have persisted at the level
measured at the end of interval 3 at least until August 19 \cite{isr98}.

The input spectral temperature $kT_0$ is consistent with a constant value of
$\approx 1$~keV during the entire observation.  The decrease in flux
following interval 1 is associated with a marginally significant decrease in
the fitted values of the scattering optical depth $\tau_P$ and the
Comptonised component normalisation parameter $A_{\rm C}$.

The model parameters associated with the gaussian component, representing
fluorescence from iron in the circumstellar matter, are consistent with
constant values over the course of the observation.  The iron line centre
energy is consistent with emission from cool matter, with no significant
change in the centre energy found between intervals.  The iron line
equivalent width (EW) increases with marginal significance following
interval 1.

\section{Pulse-phase spectroscopy}

The data from interval 1 were divided into 10 equal phase bins, and a
spectrum obtained for each phase range. The ephemeris is that of Giles et
al.  \shortcite{gil99}, with best-fit constant barycentre corrected period
$P=124.36568\pm0.00020$~s.
The primary minimum is defined as phase zero.   Data from interval 1 alone
were used, for two reasons.  Firstly, the countrate was at its highest
during that time, making the signal-to-noise ratio optimal compared to the
other intervals. It was not possible to fit models reliably to pulse-phase
spectra from interval 2 (due to the low countrate) or interval 3 (due to its
short duration). Secondly, since the evidence of the pulse profiles suggests
conditions in the source may be rather different between intervals 1, 2 and
3, this seems a better choice than simply combining all the data.

\begin{figure}
  \epsfxsize=9.0cm
  \epsfbox{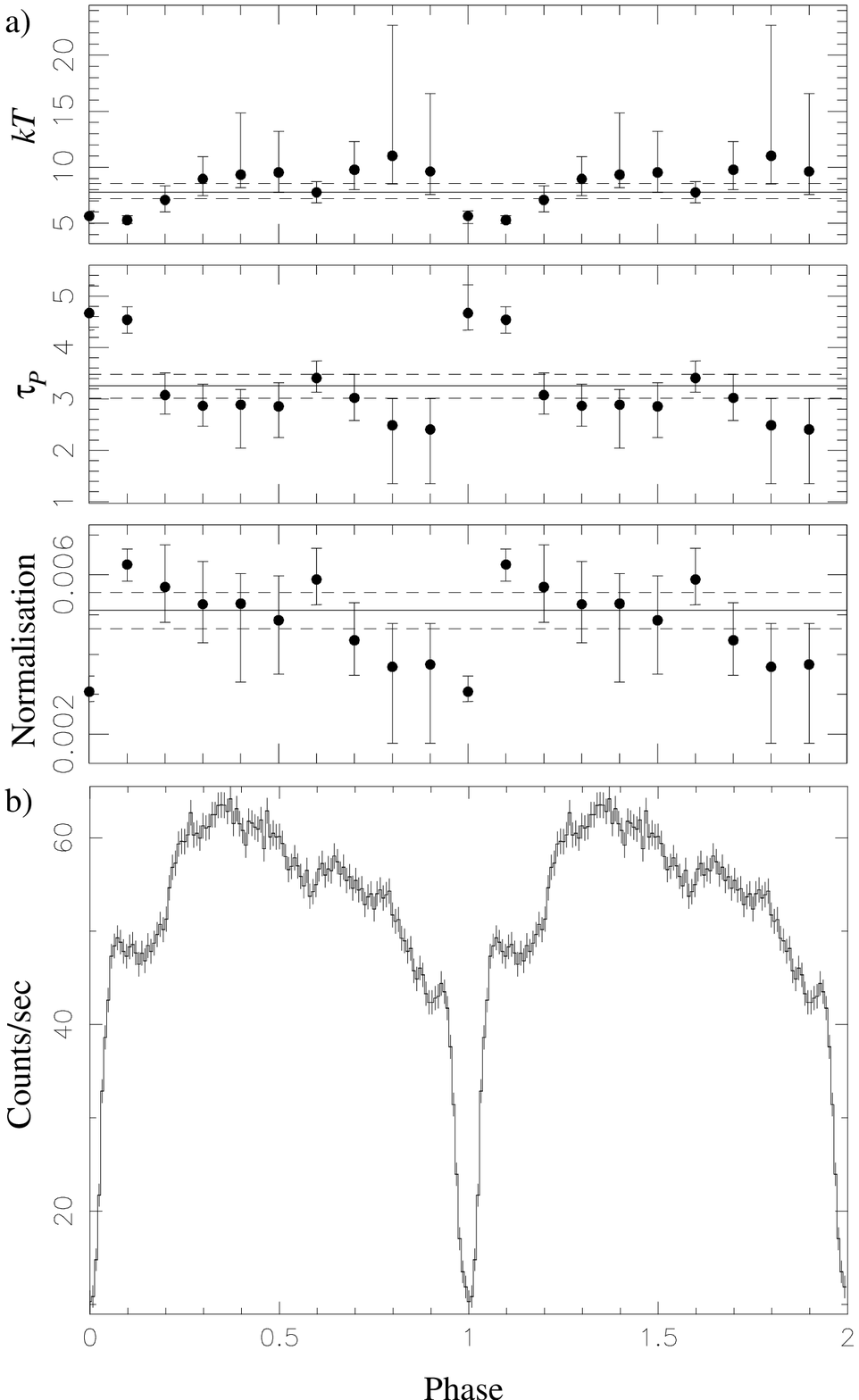}
  \caption{
a) Spectral model fit parameters with pulse phase during interval 1.  The top
panel shows the scattering plasma temperature $kT$; the middle panel the
optical depth for scattering $\tau_P$; while the bottom panel shows the model
component normalisation $A_{\rm C}$.  The errorbars show the 90 per cent
confidence limits.  The fitted parameter values for the mean interval 1
spectra are shown by the solid lines; 90 per cent confidence intervals as
dotted lines.
b)  Pulse profile from background-subtracted Event-mode PCA data. Two
full pulse periods are shown for clarity.
\label{fig2}
}
\end{figure}

Each of the 10 spectra were then fitted with the model described in section
\ref{spec}.  Initially, all fit parameters (barring those of the galactic
ridge component) were left free to vary. Fitted values of the column density
$n_H$, input spectrum temperature $kT_0$, and the iron line component
parameters were all found to be consistent with those for the mean interval 1
spectrum.  Confidence limits for the scattering plasma temperature $kT$ were
very large within some phase ranges, while significant variations with phase
were observed only in the normalisation parameter $A_{\rm C}$ and scattering
optical depth $\tau_P$. To improve the confidence intervals for the latter
three parameters a second fit was performed with all other parameters frozen
at the fitted values for the mean interval 1 spectrum. The resulting fit
values are shown in Fig. \ref{fig2}.

Around the phase of primary minimum ($\phi=0.0-0.1$) the fitted value of
$\tau_P$ is significantly higher than the mean value, while $kT$ is lower.
The normalisation $A_{\rm C}$ is also significantly lower than the mean value
at $\phi=0.0$, but in the phasebin immediately following is above the
confidence limits for the mean. There is little evidence for strong spectral
variation from the mean at other phases, however we do note an almost
monotonic decrease in the normalisation $A_{\rm C}$ from $\phi=0.1$ and
throughout the pulse cycle.

\section{Flare spectra}
\label{flare}

\begin{figure}
  \epsfxsize=9.0cm
  \epsfbox{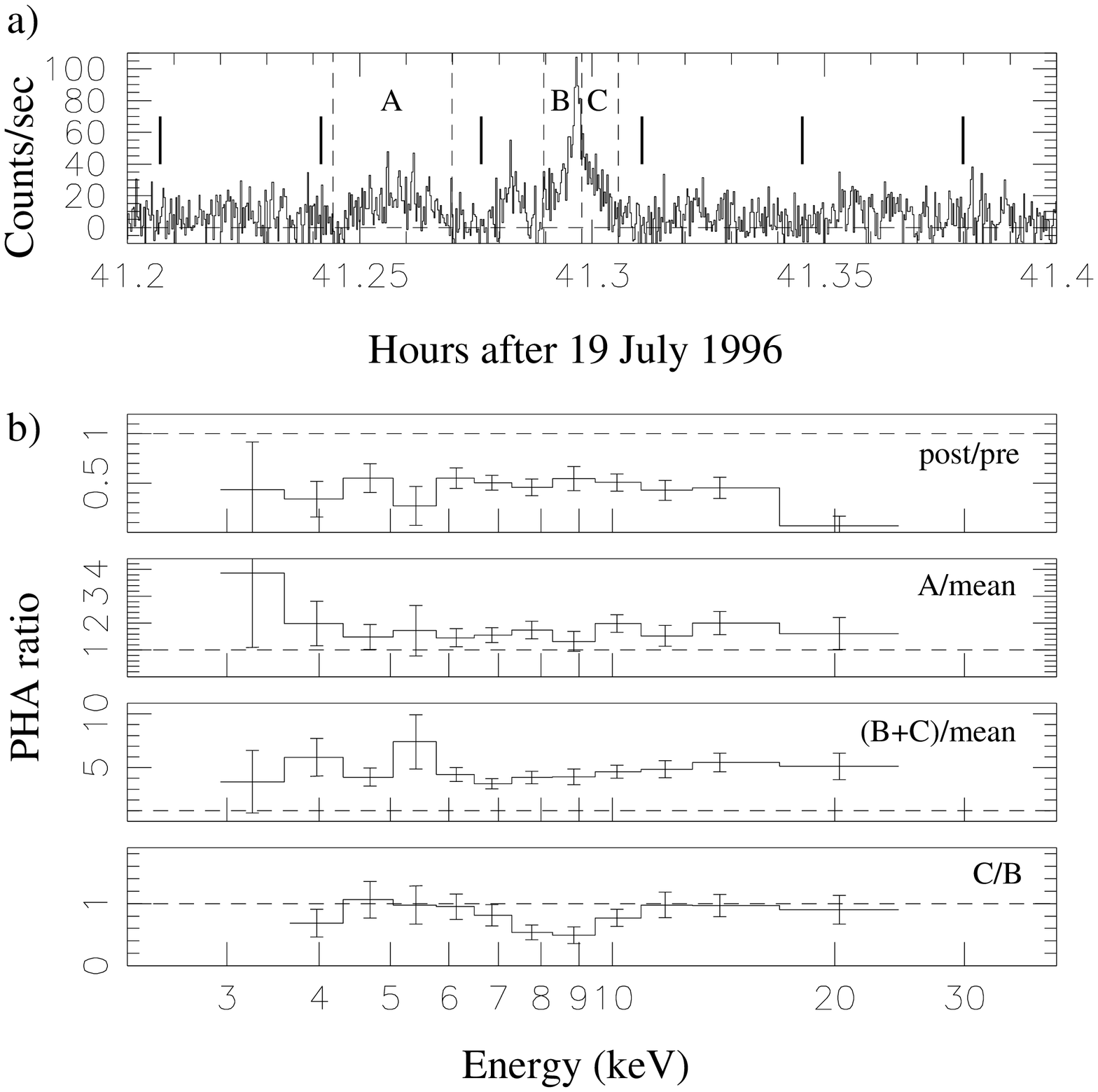}
  \caption{
a) Event-mode PCA lightcurve (averaged over 1 s bins) during interval 2
showing the flare. Thick lines show the predicted times of primary minima from
the ephemeris of Giles et al. \protect\shortcite{gil99}. The three intervals
of interest are labelled A, B and C, for the pre-flare brightening, flare
rise, and flare fall respectively.  b) The top panel shows the ratio of the
mean post-flare to the mean pre-flare spectrum (excluding the flare itself).
The second panel shows the ratio of the spectrum during the pre-flare increase
(interval A in Fig. \ref{fig3}a) to the mean spectrum (excluding the flare). The
third panel shows the ratio of the flare spectra (intervals B and C) to the mean
(excluding the flare). The fourth panel shows the ratio of the spectrum during
the flare decrease (C) to that during the flare increase (B).  
\label{fig3} }
\end{figure}

During the period of lowest flux (interval 2) a strong flare was observed,
with the peak flux rising to almost 20 times the mean level during this
interval (Fig. \ref{fig3}a).  The flare was preceeded by a modest brightening
of the source which began $\approx 150$ s before the flare itself and lasted
$\approx 60$~s; a second pre-flare brightening began $\approx 50$~s before
the main flare, lasting $\approx 30$~s. Both the flare and the pre-flare
activity occurred within the extent of two pulse periods.  From the ephemeris
determined for the full data set \cite{gil99}, a primary minimum would have
occurred between the first and second pre-flares had the source been pulsing
as was observed during intervals 1 and 3.  The instantaneous flux during the
flare peaked at $\approx 105\,{\rm count\,s^{-1}}$, compared to the mean
rate during interval 2 of $\approx 5\,{\rm count\,s^{-1}}$.  No comparable
events occurred at other times during interval 2.  During intervals 1 and 3,
the significant variations between successive pulse profiles make it
difficult to rule out flares with peaks having similar heights above the
mean level. Certainly no flares with the same proportional increase in flux
compared to the mean occurred over the course of the observation.

The count rate during this interval was too low to obtain useful
full-resolution spectra. Instead, low-resolution spectra at various times
were extracted from the uninterrupted portion of data within which the flare
was observed.  The PHA ratios obtained by dividing the various spectra are
shown in Fig. \ref{fig3}b.  The top panel shows the ratio of the mean
spectrum following the flare to that preceding it (excluding the flare
itself).   Mean flux decreased by around 50 per cent following the flare,
with no strong evidence of spectral variation. The second and third plots
show the ratios of the pre-flare (interval A on Fig. \ref{fig3}a) and flare
(intervals B and C) spectra vs. the mean (excluding the flare). In each case
there is no evidence of spectral variation; each ratio is consistent with
constant PHA ratio in the range 3-20keV. The pre-flare exhibits only a
modest increase in flux of around 50 per cent, while during the 58~s window
encompassing the flare itself the mean flux increased by 4-5 times.

The bottom panel shows the PHA ratio between the falling and rising parts of
the main flare (intervals C and B respectively). The ratio is constant
barring a broad dip between 6 and 12~keV.  Examination of the spectra
indicate that this dip is not due to any global change in the spectral
shape, but rather a localised decrease in flux within that energy range as
the flare developed.  Modelling the spectra from the rising and falling
parts using the two-temperature model described in section \ref{spec}, we
find that the variation can be fitted best by a decrease in the temperature
of the cooler component ($\chi^2_{\nu} = 1.6$) although the change in
temperature is not statistically significant.

\section{Discussion}
\label{discussion}


Re-analysed BATSE data confirm that GX~1+4 underwent a torque reversal from
spin-down to spin-up around 1996 August 2, approximately 10 days after the
RXTE observation \cite{gil99}. Since contributions to the net torque on the
neutron star may come from both accreted material and magnetic stresses
within the disc, it seems reasonable to suggest that changes in the
magnetosphere or disc structure which cause the torque reversal may occur
some time before a measurable effect is seen on the star itself.  We
therefore suggest that the spectral and pulse profile changes measured
during our observation are related to the (presently unknown) phenomenon
which causes torque reversals.  Additional support for this connection is
provided by the observation of dramatic pulse profile shape changes during
the {\it RXTE} observation coupled with the previously noted correlation
between pulse profile shape and torque state \cite{gre98}.  Until an
observation can be made which encompasses the precise time a torque reversal
is occurring it may be impossible to determine more about the process.


Comptonisation models have been used to fit spectra for this source from
past observations, e.g. Staubert et al. \shortcite{stau95}; however it has
not previously been possible to eliminate all other candidate models on the
basis of the $\chi^{2}$ fit parameter. The particular model used for the
spectral fitting simulates Comptonisation in an unmagnetised plasma
\cite{tit94}, and since the available evidence points towards a strong
magnetic field in GX~1+4 (although this awaits confirmation by more direct
measurements such as a cyclotron resonance line) the model fit parameters
may not be an accurate measure of the source conditions. It is likely that
the principal effect of the magnetic field will be to make the spectral
parameters dependent on the emission angle. Hence the model fit parameters
obtained from the mean spectra are expected to be a reasonable approximation
of the actual values (L. Titarchuk, pers.  comm.)

Assuming that the majority of the X-ray emission originates from a blackbody
at most the size of the neutron star ($R_* \approx 10$~km), we expect a
temperature $kT_0 \ga 0.5$~keV. The temperature of the input spectrum $kT_0
\approx 1$~keV obtained from the model fits is consistent with this
calculation.  Rough estimates of the accretion column density can be made
based on the mass transfer rate derived from the luminosity, and assuming a
simple column geometry. The accretion luminosity $L_{acc} \approx GM_*
\dot{M}/R_*$ and hence during interval 1 $\dot{M} \approx 2 \times
10^{16}\,{\rm g\,s^{-1}}$.  Assuming that the accretion column radius $R_c$
is some fraction $f$ of the neutron star radius $R_*$, and the column plasma
is moving at approximately the free fall velocity $\approx 0.5c$, the
estimated optical depth for Thompson scattering is $\approx 0.17/f$. In
general $f$ is subject to considerable uncertainties particularly given the
over-simplistic geometry adopted here, but we estimate $f\approx 4\times
10^{-2}$ (e.g. Frank, King and Raine 1992) and thus the optical depth
$\tau\approx5$, close to the model fit values.


The pulse phase spectroscopy results also show that $\tau_P$, $kT$ and $A_{\rm
C}$ are significantly modulated at the pulsar rotation period.  Consequently
we propose that the Comptonisation model provides a realistic picture of
spectral formation in this source, with scattering taking place in the
accretion column.  Thus the $kT$ parameter can be interpreted as the mean
temperature of the accretion column plasma over the region in the column where
scattering takes place.  The model normalisation parameter $A_{\rm C}$ is
somewhat more difficult to relate to a physically measurable quantity, since
both the $kT$ and $\tau_P$ parameters can also affect the total flux from the
model component.


The spectral ratios (Fig. \ref{fig1}c) and the spectral fit parameters
strongly suggest that the variations in the mean spectra during the course
of the observation are due to two factors. The decrease in flux which is
essentially independent of energy is presumably a result of decreased rate
of mass transfer to the neutron star ${\dot M}$.  This is accompanied by a
strong increase in absorption by cold material causing the flux decrease
below 6~keV.


Variations in the column density $n_H$ on time-scales of $\approx 2$~h have
not previously been observed in this source. The iron line energy and the
relationship between equivalent width and $n_H$ are consistent with the
spherical distribution of matter suggested by Kotani et al.
\shortcite{kot99}, however the variation is much too rapid to be
attributable to the negative feedback effect which those authors suggest
regulates mass transfer to the neutron star in the long term.  The rapid
variation may be an indication of significant inhomogeneities in the
circumstellar matter, or alternatively that the giant wind velocity is much
faster than 10~${\rm km\,s}^{-1}$ as suggested by infrared observations of
the companion \cite{chak98}. 


Variation in the spectral fit parameters with pulse phase may provide clues
to the distribution of matter in the accretion column. The sharp dip in the
pulse profiles is associated with a significant increase in the scattering
optical depth $\tau_P$ and decrease in the Comptonisation component
normalisation parameter $A_{\rm C}$ (Fig. \ref{fig2}). Such an effect may be
observed if the accretion column is viewed almost directly along the
magnetic axis, resulting in a much greater path length for photons
propagating through the relatively dense matter of the column; essentially
an `eclipse' of the neutron star pole by the accretion column.  Preliminary
Monte Carlo modelling based on Comptonisation as the source of high-energy
photons supports this as a possible mechanism (Galloway, 1999, work in
progress).  Accretion column eclipses have previously been postulated to
explain dips in pulse profiles from A~0535+262 \cite{cem98} and
RX~J0812.4-3114 \cite{rei99}.  That the plasma temperature $kT$ is also low
around the phase of primary minimum may be related to the bulk motion of the
column plasma, since the relative velocity of the plasma in the observer's
frame will depend on orientation (and hence pulse phase). The velocity of
bulk motion is likely to be many orders of magnitude above the thermal
velocity (in the plasma rest frame) and so may result in observable
variation of this fit parameter with pulse phase. The asymmetry of the
normalisation $A_{\rm C}$ with respect to the primary minimum furthermore
points to significant asymmetry of the emission on the `leading' and
`trailing' side of the pole.  Such asymmetry may originate from a nonzero
relative velocity between the disc and column plasma where the disc plasma
becomes bound to the magnetic field lines and enters the magnetosphere
\cite{waw81}.  The additional observation that the width of the dip
decreases with increasing energy may point towards a role for resonant
absorbtion \cite{gil99}.


The observation of a short-duration flare during the minimum flux period
provides a further example of previously unseen behaviour in this source.
With the peak flux during the flare rising to almost 20 times the mean
level, and with no other comparable events observed during interval 2, it is
likely that the flare was due to a short-lived episode of enhanced
accretion. The mean accretion rate during interval 2 can be estimated to be
$\approx 2.5 \times 10^{15}\,{\rm g\,s^{-1}}$ \cite{fra92}; the increased
luminosity observed during the flare thus implies additional accretion of at
least $5\times 10^{17}$~g. In order to measure the instantaneous $\dot{M}$
throughout the flare it would be necessary to correct for the effects of
anisotropic emission from the neutron star surface as well as changing
observation angle with the star's rotation. Since the geometry is essentially
unknown, and beam patterns rather model dependent, this is not yet possible.

We do however note that the delay between the start of the pre-flare
increase (`A' in Fig. \ref{fig3}) and the flare itself is $\approx 150$~s.
The relative angular velocities of the disc plasma and the neutron star
magnetic field lines at the inner disc radius imply a periodicity
significantly different from that resulting from the neutron star's
rotation.  From the mean interval 2 luminosity and the estimated surface
magnetic field strength for GX~1+4 of $3\times 10^{13}$~G we estimate the
inner disc radius as $2.7 \times 10^{7}$~m. A locally dense patch of plasma
rotating with Keplerian velocity in the disc would pass close to the region
where plasma enters the accretion column originating from each pole every
150~s or so. Thus it is conceivable that these two events represent
successive passages of the same patch through the column uptake zone in the
disc. After the second passage, the patch is presumably completely
transmitted to the star and so no further flaring behaviour is seen.  

If ${\dot M}$ variations are a significant factor in the evolution of the
flare, we might see other indications in the flare shape. If the polar
region cools much more slowly than the flare timescale an asymmetric flare
might be observed.  Spectral model fits might also indicate cooling of the
emission component originating from the pole. However, the flare appears
almost completely symmetric, and spectral fits to the rising and falling
parts of the flare do not exhibit cooling at any statistically significant
level.

\section*{Acknowledgments}

We would like to thank Dr. K. Wu for many helpful discussions and
suggestions during the preparation of this paper.  The {\it RXTE} GOF
provided timely and vital help and information, as well as the archival
observations from 1996 and 1997. We would also like to thank the BATSE
pulsar group for providing the timing data.

\label{lastpage}

\end{document}